\documentclass[12pt]{article}
\usepackage{amsmath}
\usepackage{amssymb}
\usepackage{amsthm}
\usepackage{xcolor}
\usepackage{caption}
\usepackage{calc}
\usepackage{hyperref}
\usepackage{graphicx}
\usepackage{wrapfig}
\usepackage{enumitem}
\usepackage{cite}
\usepackage{geometry}
\usepackage{algorithm}
\usepackage[noend]{algpseudocode}
\newcounter{algsubstate}
\renewcommand{\thealgsubstate}{\alph{algsubstate}}

\begin{document}

\title{Eigenvector centrality for multilayer networks with dependent node importance}

\author{H. Robert Frost$^{1}$}
\date{}
\maketitle
\begin{center}
\textit{
$^1$Department of Biomedical Data Science\\
Geisel School of Medicine \\
Dartmouth College \\
Hanover, NH 03755, USA \\
rob.frost@dartmouth.edu
}
\end{center}

\begin{abstract}
We present a novel approach for computing a variant of eigenvector centrality for multilayer networks with inter-layer constraints on node importance. Specifically, we consider a multilayer network defined by multiple edge-weighted, potentially directed, graphs over the same set of nodes with each graph representing one layer of the network and no inter-layer edges. As in the standard eigenvector centrality construction, the importance of each node in a given layer is based on the weighted sum of the importance of adjacent nodes in that same layer. Unlike standard eigenvector centrality, we assume that the adjacency relationship and the importance of adjacent nodes may be based on distinct layers. Importantly, this type of centrality constraint is only partially supported by existing frameworks for multilayer eigenvector centrality that use edges between nodes in different layers to capture inter-layer dependencies. For our model, constrained, layer-specific eigenvector centrality values are defined by a system of independent eigenvalue problems and dependent pseudo-eigenvalue problems, whose solution can be efficiently realized using an interleaved power iteration algorithm. 
\end{abstract}

\section{Eigenvector centrality for multilayer networks}\label{sec:background}

Computation of node importance via centrality measures is an important task in network analysis and a large number of centrality measures have been developed that prioritize different node and network properties \cite{Newman:2010ve}. The most widely used centrality measures are a function of the network adjacency matrix, $\mathbf{A}$, which, for an edge-weighted network defined over $p$ nodes is the $p \times p$ matrix:

\begin{align} \label{eqn:adj_matrix}
\mathbf{A} &= \begin{bmatrix}
a_{1,1} & \cdots & a_{1,p} \\
\vdots & \ddots & \vdots \\
a_{p,1} & \cdots & a_{p,p}
\end{bmatrix}
\end{align}

\noindent where $a_{i,j}$ captures the weight of the edge between nodes $i$ and $j$ or 0 if no edge exists between these nodes. Self-edges are represented by elements on the diagonal. If the network is directed, then $a_{i,j}$ and $a_{j,i}$ capture distinct edges and $\mathbf{A}$ is asymmetric; if the network is undirected, $a_{i,j} = a_{j,i}$ and $\mathbf{A}$ is symmetric. 

Modeling node importance as the weighted sum of the importance of adjacent nodes leads a version of centrality called eigenvector centrality, which is solved by computing the principal eigenvector of the following eigenvalue problem:

\begin{align} \label{eqn:standard_ec}
\mathbf{A} \mathbf{x} &= \lambda \mathbf{x}
\end{align}

\noindent Specifically, the eigenvector centrality for node $n$ is given by element $n$ of the principal eigenvector $\mathbf{x}$ corresponding to the largest eigenvalue \cite{Newman:2010ve}. When $\mathbf{A}$ is irreducible (i.e., the network is strongly connected), then the Perron-Frobenius theorem \cite{Perron:1907ua} guarantees that there is a unique largest real eigenvalue whose corresponding eigenvector can be chosen to have strictly positive elements. For directed graphs, left and right versions of eigenvector centrality are possible, i.e., the solution to the eigenvalue problem for $\mathbf{A}^T$ or $\mathbf{A}$. For the methods developed below, we focus on the right eigenvector centrality, however, the same approach can be employed to compute left eigenvector centrality by considering $\mathbf{A}^T$  instead of $\mathbf{A}$.

In this paper, we are interested in eigenvector centrality and how that measure of node importance generalizes to multilayer (or multiplex) networks \cite{10.1093/comnet/cnu016, PhysRevE.89.032804}. We assume a multilayer network comprised by $k$ layers that each represent a potentially directed, edge-weighted graph over the same $p$ nodes. The graph for layer $j \in \{1,...,k\}$ can be represented by the adjacency matrix $\mathbf{A}_j$:
\begin{align} \label{eqn:multilayer_adjacency}
\mathbf{A}_j &= \begin{bmatrix}
a_{j,1,1} & \cdots & a_{j,1,p} \\
\vdots & \ddots & \vdots \\
a_{j,p,1} & \cdots & a_{j,p,p}
\end{bmatrix}
\end{align}
\noindent where $a_{j,n,m}$ holds the weight of the edge from node $n$ to node $m$ within the layer $j$ graph. Although the terms network and graph are synonymous in this context, we will generally use the term network to refer to the entire multilayer network and the term graph to refer to the network that defines a single layer.
In the context of a multilayer network, node eigenvector centrality can be evaluated at the level of a specific layer (i.e., a given node has separate centrality values for each of the $k$ layers) or at the level of the entire multilayer network (i.e., a given node has a single centrality value that captures the importance of the node across all $k$ layers). In the development below, we focus on layer-specific measures of eigenvector centrality.

If the $k$ layers are independent, then eigenvector centrality can simply be computed separately for each layer. However, if dependencies exist between the layers, then a multilayer version of eigenvector centrality must be employed that can account for the inter-layer constraints. A number of approaches for modeling and computing multilayer eigenvector centrality have been explored over the last decade (e.g., \cite{doi:10.1137/17M1137668, doi:10.1137/19M1262632, De-Domenico:2015wa, DBLP:journals/compnet/DeFordP18, doi:10.1063/1.4818544}).
Most of these approaches assume that inter-layer constraints can be modeled by edges between the nodes in one layer and nodes in other layers. This type of approach is exemplified by the recent work of Taylor et al. \cite{doi:10.1137/19M1262632} that details a flexible model for a "uniformly and diagonally coupled multiplex network".
Specifically, Taylor et al. represent inter-layer dependencies by equally weighted edges connecting the nodes in one layer to the same nodes in a dependent layer. Taylor et al. represent the structure of these dependencies using a $k \times k$ adjacency matrix $\tilde{\mathbf{A}}$:
\begin{align} \label{eqn:layer_adjanency}
\tilde{\mathbf{A}} &= \begin{bmatrix}
\tilde{a}_{1,1} & \cdots & \tilde{a}_{1,k} \\
\vdots & \ddots & \vdots \\
\tilde{a}_{k,k} & \cdots & \tilde{a}_{k,k}
\end{bmatrix}
\end{align}
\noindent where $\tilde{a}_{i,j}$ represents the weight of the edges from nodes in layer $i$ to nodes in layer $j$. Computation of multilayer eigenvector centralities is then based on the principal eigenvector of a $kp \times kp$ supercentrality matrix $\mathbb{C}(\omega)$:

\begin{align} \label{eqn:supercentrality}
\mathbb{C}(\omega) &= \hat{\mathbb{C}} + \omega \hat{\mathbb{A}}
\end{align}

\noindent where $\hat{\mathbb{C}} = diag[\mathbf{A}_1, ... , \mathbf{A}_k]$ (i.e., a $kp \times kp$ block diagonal matrix that has the adjacency matrices for each of the $k$ layers along the diagonal), $\hat{\mathbb{A}} = \tilde{\mathbf{A}} \otimes \mathbf{I}$ (i.e., the Kronecker product of $\tilde{\mathbf{A}}$ and $\mathbf{I}$), and $\omega$ is the coupling strength. The principal eigenvector of $\mathbb{C}(\omega)$ can then be used to find joint, marginal, and conditional eigenvector centralities. Specifically, the principal eigenvector elements are divided into $k$ sequential blocks of $p$ elements, with the block corresponding to layer $i$ representing the joint centrality values for the nodes in layer $i$. To calculate the marginal centralities for either nodes or layers, the joint centralities are summed over all layers for a given node or all nodes for a given layer. To calculate conditional centralities for either nodes or layers, the joint centrality value for a given node/layer pair is divided by either the marginal centrality for the layer or the marginal centrality for the node. 

\section{Eigenvector centrality for multilayer networks with inter-layer constraints on adjacent node importance}\label{sec:network_model}

Our model for multilayer eigenvector centrality extends the standard single network version given by \eqref{eqn:standard_ec} to support the scenario where the importance of a given node in layer $i$ is proportional to the weighted sum of the importance of adjacent nodes with adjacency and weights based on layer $i$ but adjacent node importance based on potentially distinct layers. This model is conceptually and mathematically distinct from approaches like Taylor et al. that add edges between the same node in different layers to capture inter-layer dependencies. 
To illustrate, assume we have a multilayer network with just two layers $i$ and $j$. If we assume the importance for nodes in layer $i$ has the standard definition, i.e., it is not dependent on another layer, the solution is given by the typical eigenvalue problem:

\begin{align} \label{eqn:indep_example}
\mathbf{A}_i \mathbf{x}_i &= \lambda_i \mathbf{x}_i
\end{align}

However, if we assume the importance for nodes in layer $j$ is based on the importance of adjacent nodes in layer $i$, then solution for layer $j$ is given by the following linear model (note that both $\mathbf{x}_i$ and $\mathbf{x}_j$ are included):

\begin{align} \label{eqn:dep_example}
\mathbf{A}_j \mathbf{x}_i &= \lambda_j \mathbf{x}_j
\end{align}

\noindent Although the linear model \eqref{eqn:dep_example} is not technically an eigenvalue problem since it contains distinct $\mathbf{x}_i$ and $\mathbf{x}_j$ vectors, we will refer to it as a pseudo-eigenvalue problem given the structural similarities to \eqref{eqn:indep_example} and the fact that $\mathbf{x}_i$ may represent an eigenvector.  As detailed below, we will also use the pseudo-eigenvalue label to describe more complex scenarios where $\mathbf{x}_i$ found on both sides of the equation. For this example, the solution for the entire multilayer network is given by a system of an independent eigenvalue problem and a dependent pseudo-eigenvalue problem:

\begin{align} \label{eqn:example_system}
\begin{split}
\mathbf{A}_i \mathbf{x}_i &= \lambda_i \mathbf{x}_i \\
\mathbf{A}_j \mathbf{x}_i &= \lambda_j \mathbf{x}_j
\end{split}
\end{align}

\noindent In this case, the solution can be obtained by first solving the eigenvalue problem for layer $i$ to find $\mathbf{x}_i$ and then computing $\mathbf{x}_j$ as $\mathbf{x}_j = 1/\lambda_j \mathbf{A}_j \mathbf{x}_i$ with the value of $\lambda_j$ set to ensure $\mathbf{x}_j$ is unit length. 

%
%

This simple example can be generalized to a multilayer network with $k$ layers and arbitrary node importance constraints encoded by a graph whose nodes represent layers and whose weighted and directed edges represent inter-layer dependencies. Let this inter-layer dependency graph be represented by the $k \times k$ adjacency matrix $\tilde{\mathbf{A}}$ that is similar in structure to the $\tilde{\mathbf{A}}$ used by Taylor et al. and defined in \eqref{eqn:layer_adjanency} but with the added constraint that the rows must sum to 1 (i.e., $\forall_{i \in 1,...,k} \sum_{j=1}^{k} \tilde{a}_{i,j} = 1$). Element $\tilde{a}_{i,j}$ of $\tilde{\mathbf{A}}$ represents the strength of the dependency between adjacent node importance in layer $i$ and node importance in layer $j$ with the sum of all dependencies for a given layer equal to 1.  If $\tilde{\mathbf{A}} = \mathbf{I}$, all of the layers are independent. For the 2 layer example represented by \eqref{eqn:example_system}, $\tilde{\mathbf{A}}$ is:
\begin{align} \label{eqn:example_adjanency}
\tilde{\mathbf{A}} &= \begin{bmatrix}
1 & 0 \\
1 & 0 
\end{bmatrix}
\end{align}

\noindent If the length $p$ vector $\mathbf{x}_i$ represents node importance in layer $i$, we can define an adjacent node importance function $\mathbf{c}(i,\tilde{\mathbf{A}})$ as:

\begin{align} \label{eqn:dependency_function}
\mathbf{c}(i, \tilde{\mathbf{A}}) &= \sum_{j=1}^{k} \tilde{a}_{i,j} \mathbf{x}_j
\end{align}

\noindent In other words, the importance of nodes in layer $i$ is based on a weighted sum of the importance of adjacent nodes in other layers (note that adjacency is only based on the topology of layer $i$). Given the function $\mathbf{c}()$, we can compute a constrained multilayer version of eigenvector centrality for the network by solving the following system of $k$ interdependent eigenvalue and pseudo-eigenvalue problems:

\begin{align}\label{eqn:eigen_system}
\begin{split}
\mathbf{A}_1 \mathbf{c}(1, \tilde{\mathbf{A}} ) &= \lambda_1 \mathbf{x}_1 \\
\mathbf{A}_2 \mathbf{c}(2, \tilde{\mathbf{A}})  &= \lambda_2 \mathbf{x}_2 \\
&\vdots \\
\mathbf{A}_k \mathbf{c}(k, \tilde{\mathbf{A}}) &= \lambda_k \mathbf{x}_k \\
\end{split}
\end{align}

\noindent Importantly, the supercentrality approach of Taylor et al. cannot in general solve systems such as \eqref{eqn:example_system}, i.e., system \eqref{eqn:eigen_system} cannot be directly mapped to an eigenvalue problem involving a supercentrality matrix of the form defined by \eqref{eqn:supercentrality}.

In the special case where $\tilde{\mathbf{A}} = \mathbf{I}$, all $\mathbf{c}(i, \tilde{\mathbf{A}}) = \mathbf{x}_i$ and \eqref{eqn:eigen_system} becomes a system of $k$ independent eigenvalue problems:

\begin{align}\label{eqn:identity_case}
\begin{split}
\mathbf{A}_1 \mathbf{x}_1 &= \lambda_1 \mathbf{x}_1 \\
\mathbf{A}_2 \mathbf{x}_2  &= \lambda_2 \mathbf{x}_2 \\
&\vdots \\
\mathbf{A}_k \mathbf{x}_k &= \lambda_k \mathbf{x}_k \\
\end{split}
\end{align}

\noindent More generally, the dependency structure for a given layer $i$ falls into one of three cases:

\begin{enumerate}[label=\Alph*]
\item $\tilde{a}_{i,i} = 1$.  In this scenario, the eigenvector centrality for layer $i$ is given by the principal eigenvector of the independent eigenvalue problem $\mathbf{A}_i \mathbf{x}_i = \lambda_i \mathbf{x}_i$.
\item $\tilde{a}_{i,i} = 0$: In this scenario, the centrality for layer $i$ is a linear function of the centrality values of other network layers.
\item $0 < \tilde{a}_{i,i} < 1$: In this scenario, the centrality for layer $i$ is given by a pseudo-eigenvalue problem that can be rewritten as $\mathbf{A}_i \mathbf{x}_i + \mathbf{d} = \lambda_i \mathbf{x}_i$, where $\mathbf{d}$ is captures the part of $\mathbf{A}_i \mathbf{c}(i, \tilde{\mathbf{A}} )$ not due to $\mathbf{x}_i$. 
\end{enumerate}

%
%

\noindent If all layers in the multilayer network fall into case A or B and no cycles exist in the graph defined by $\tilde{\mathbf{A}}$, then the constrained eigenvector centralities can be computed using a relatively straightforward two-step procedure:

\begin{enumerate}
\item Solve the independent eigenvalue problems for all layers in case A using an algorithm like power iteration  \cite{MisesPraktischeVD}.
\item Sequentially solve the linear models for all layers in case B with the order of solution given by the inter-layer constraints.
\end{enumerate}

\noindent If any layers fall into case C or cycles exist in the inter-layer dependency graph, then the solution must be obtained via an iterative algorithm similar to the interleaved power iteration approach detailed in Section \ref{sec:algorithm} as Algorithm 1.


\section{Interleaved power iteration algorithm for a system of dependent pseudo-eigenvalue problems}\label{sec:algorithm}


For an arbitrary inter-layer dependency matrix $\tilde{\mathbf{A}}$, the joint solution for system \eqref{eqn:eigen_system} can be found via a interleaved version of the power iteration method, detailed in Algorithm 1, that is applied across all $k$ linear problems.
It should be noted that this specification of Algorithm 1 does not include features important for many practical implementations, e.g, checks to ensure the input matrices $\mathbf{X}_i$ are well conditioned, options for the use of stochastic initialization of the eigenvectors, use of techniques like accelerated stochastic power iteration \cite{pmlr-v84-xu18a} to improve computational performance, alternate stopping conditions, etc..

\begin{algorithm}\label{alg:eespca}
\caption{Interleaved power iteration for dependent pseudo-eigenvalue problems}
\hspace*{\algorithmicindent} \textbf{Input:} 
\begin{itemize}
\setlength\itemsep{0em}
\item Set of $k$ $p \times p$ irreducible matrices, $\{\mathbf{X}_1, \mathbf{X}_2, ..., \mathbf{X}_k\}$
\item Dependencies between the principal pseudo-eigenvectors of $\mathbf{X}_i$ encoded as a $k \times k$ matrix $\tilde{\mathbf{A}}$ whose rows sum to 1 (see \eqref{eqn:eigen_system})
The pseudo-eigenvalue problem for $\mathbf{X}_i$ is given by $\mathbf{X}_i \mathbf{c}(i, \tilde{\mathbf{A}}) = \lambda_i \mathbf{x}_i$, where $\mathbf{c}(i, \tilde{\mathbf{A}}) = \sum_{j=1}^{k} \tilde{a}_{i,j} \mathbf{x}_j$
\item Positive integer $maxIter$ that represents the maximum number of iterations
\item Positive real number $tol$ that represents the stopping criteria as the proportional change in the mean of the $k$ pseudo-eigenvalues between iterations
\end{itemize}
\hspace*{\algorithmicindent} \textbf{Output:} 
\begin{itemize}
\setlength\itemsep{0em}
\item Estimated principal pseudo-eigenvectors of the input $\mathbf{X}_i$: $\{\hat{\mathbf{v}}_1, \hat{\mathbf{v}}_2, ..., \hat{\mathbf{v}}_k\}$
\item Estimated principal pseudo-eigenvalues of the input $\mathbf{X}_i$: $\{\hat{\lambda}_1, \hat{\lambda}_2, ..., \hat{\lambda}_k\}$
\item Number of iterations completed
\end{itemize}
\hspace*{\algorithmicindent} \textbf{Notation:} 
\begin{itemize}
\setlength\itemsep{0em}
\item Let $\mathbf{v}_{n,m}$ and $\lambda_{n,m}$ represent the principal pseudo-eigenvector and pseudo-eigenvalue for matrix $\mathbf{X}_n$ as computed on the $m^{th}$ iteration of the algorithm
\end{itemize}
\begin{algorithmic}[1]
\State $\forall_{j \in 1...k }\mathbf{v}_{j,0} = \{1/\sqrt{p}, ... , 1/\sqrt{p}\}$
	\Comment{Initialize principal pseudo-eigenvectors to unit length vectors with all values equal to $1/\sqrt{p}$}
\For{$i \in \{1,...,maxIter\}$}
\State $\forall_{j \in 1...k } \mathbf{v}_{j,i} = \mathbf{X}_j \mathbf{c}(i, \tilde{\mathbf{A}})$
 \Comment{Update principal pseudo-eigenvectors based on dependencies}

\State $\forall_{j \in 1...k } \mathbf{v}_{j,i} = \mathbf{v}_{j,i}/||\mathbf{v}_{j,i}||$
 \Comment{Normalize pseudo-eigenvectors to unit length}
 
\State $\forall_{j \in 1...k } \lambda_{j,i} = (\mathbf{v}_{j,i})^T \mathbf{X}_j \mathbf{c}(i, \tilde{\mathbf{A}})$
 \Comment{Update principal pseudo-eigenvalues}
  
\If{$i > 1$}
  \State $\Delta = (1/k\sum_{j=1}^{k} | \lambda_{j, i-1} - \lambda_{j,i}| )/( 1/k \sum_{j=1}^{k} \lambda_{j,i})$ 
   \Comment{Compute proportional change in mean pseudo-eigenvalue}
  \If{$\Delta < tol$}
    \State \textbf{break} \Comment{If proportion change is less than $tol$, exit}
  \EndIf
\EndIf
\EndFor
\Return {$\{\mathbf{v}_{1,i}, \mathbf{v}_{2,i}, ..., \mathbf{v}_{k,i}\}, \{\lambda_{1,i}, \lambda_{2,i}, ..., \lambda_{k,i}\}, i$}

\end{algorithmic}
\end{algorithm}

\clearpage

\section{Example problem}

\begin{figure}[t]
\begin{center}
\includegraphics[width=0.6\textwidth]{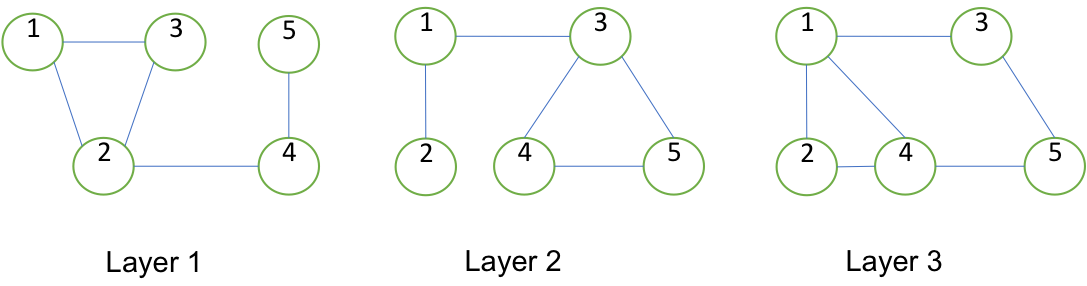}
\end{center}
\caption{Example undirected and unweighted multilayer network.}
\label{fig:example}
\end{figure}

To illustrate the constrained multilayer model detailed in Section \ref{sec:network_model} and the performance of the interleaved power iteration algorithm detailed in Section \ref{sec:algorithm}, we consider a simple multilayer network comprised by three layers that each define an undirected and non-weighted network with five nodes. The structure of this multilayer network is shown in Figure \ref{fig:example}.
For this example network, the symmetric adjacent matrices for the two layers are given by:

\begin{align*} \label{eqn:example_adjacency}
\begin{split}
\mathbf{A}_1 &= \begin{bmatrix}
0 & 1 & 1 & 0 & 0 \\
1 & 0 & 1 & 1 & 0 \\
1 & 1 & 0 & 0 & 0 \\
0 & 1 & 0 & 0 & 1 \\
0 & 0 & 0 & 1 & 0 
\end{bmatrix},
\mathbf{A}_2 = \begin{bmatrix}
0 & 1 & 1 & 0 & 0 \\
1 & 0 & 0 & 0 & 0 \\
1 & 0 & 0 & 1 & 1 \\
0 & 0 & 1 & 0 & 1 \\
0 & 0 & 1 & 1 & 0 
\end{bmatrix},
\mathbf{A}_3 = \begin{bmatrix}
0 & 1 & 1 & 1 & 0 \\
1 & 0 & 0 & 1 & 0 \\
1 & 0 & 0 & 0 & 1 \\
1 & 1 & 0 & 0 & 1 \\
0 & 0 & 1 & 1 & 0 
\end{bmatrix} \\
\end{split}
\end{align*}

\noindent We consider four different inter-layer dependency scenarios:

\begin{enumerate}
\item \textbf{No dependencies}

If no dependencies exist between the layers (i.e., $\tilde{\mathbf{A}}= \mathbf{I}$), the eigenvector centralities (rounded to two decimal places) for each layer are:

\begin{align*}
\begin{split}
\mathbf{v}_1 &= \{0.50,0.60,0.50,0.34,0.15\}\\
\mathbf{v}_2 &= \{0.34,0.15,0.60,0.50,0.50\} \\
\mathbf{v}_3 &= \{0.53, 0.43, 0.36, 0.53, 0.36\}
\end{split}
\end{align*}

As expected given the structure of layer 1, node 2 has the largest eigenvector centrality, followed by nodes 1 and 3 with node 5 having the lowest. Similarly for layer 2, node 3 has the largest centrality, followed by nodes 4 and 5 with node 2 having the lowest centrality. For layer 3, nodes 1 and 4 are tied for the largest centrality with nodes 3 and 5 tied for the lowest.

\item \textbf{Mixture of layer dependency cases A and B}

If layer 1 is independent, layer 2 is dependent on just layer 1 and layer 3 is dependent on layer 2, the $\tilde{\mathbf{A}}$ matrix takes the form:

\begin{align*}
\tilde{\mathbf{A}} = \begin{bmatrix}
1 & 0 & 0 \\
1 & 0 & 0 \\
0 & 1 & 0 
\end{bmatrix} 
\end{align*}

\noindent and the constrained eigenvector centralities are:
\begin{align*}
\begin{split}
\mathbf{v}_1 &= \{0.50,0.60,0.50,0.34,0.15\}\\
\mathbf{v}_2 &= \{0.58,0.26,0.53,0.34,0.44\} \\
\mathbf{v}_3 &= \{0.48, 0.39, 0.43, 0.54, 0.37\}
\end{split}
\end{align*}

\noindent Since layer 1 is still independent in this scenario, it has the same centrality values as the prior case. 
For layer 2, we see the expected increase in the centrality of node 1 relative to node 3 given the importance of their adjacent nodes in layer 1 (i.e, node 1 is adjacent to node 2, which has the largest centrality value in layer 1; node 3 is adjacent to nodes 4 and 5, which have the lowest centrality values in layer 1). For layer 3, the centrality for node 3 is has the largest change (an increase) relative to the independent scenario, which is expected given that it is adjacent to the node with the largest centrality value in layer 2 (node 1).

\item \textbf{Mixture of layer dependency cases A, B and C}

If layer 1 is independent, layer 2 is dependent on just layer 1 and layer 3 is equally dependent on both layer 2 and itself, the $\tilde{\mathbf{A}}$ matrix takes the form:

\begin{align*}
\tilde{\mathbf{A}} = \begin{bmatrix}
1 & 0 & 0 \\
1 & 0 & 0 \\
0 & 0.5 & 0.5 
\end{bmatrix} 
\end{align*}

\noindent and the constrained eigenvector centralities are:
\begin{align*}
\begin{split}
\mathbf{v}_1 &= \{0.50,0.60,0.50,0.34,0.15\}\\
\mathbf{v}_2 &= \{0.58,0.26,0.53,0.34,0.44\} \\
\mathbf{v}_3 &= \{0.51,0.41,0.39,0.53,0.37\}
\end{split}
\end{align*}

\noindent Since layers 1 and 2 have the same dependency structure as the prior scenario, the centrality values are unchanged. As expected, the equally divided dependency structure for layer 3 yields centrality values that are between those computed in the first two scenarios.

\item \textbf{All layers are dependency case B with cycle}

If layer 1 is dependent on layer 3, layer 2 dependent on layer 1 and layer 3 dependent on layer 2, a cycle is introduced in the layer dependency graph, the $\tilde{\mathbf{A}}$ matrix takes the form:

\begin{align*}
\tilde{\mathbf{A}} = \begin{bmatrix}
0 & 0 & 1 \\
1 & 0 & 0 \\
0 & 1 & 0 
\end{bmatrix} 
\end{align*}

\noindent and the constrained eigenvector centralities are:
\begin{align*}
\begin{split}
\mathbf{v}_1 &= \{0.40,0.68,0.42,0.38,0.25\}\\
\mathbf{v}_2 &= \{0.58,0.21,0.55,0.36,0.43\} \\
\mathbf{v}_3 &= \{0.48,0.40,0.43,0.52,0.39\}
\end{split}
\end{align*}

\end{enumerate}

\section{Applications and future directions}

We believe the inter-layer dependency model outlined in this paper has utility for a number of real world multilayer network analysis problems where node adjacency and adjacent node importance are captured by distinct networks.
One specific example of such a real world problem involves the characterization of ligand/receptor mediated cell-cell communication within a tissue. This cell signaling problem was in fact the original motivation for our method.  A simplistic model for this problem uses a fully connected network whose nodes represent cells and with edge weights based on the inverse squared Euclidean distance between each cell to capture secreted protein diffusion. In this scenario, we assume that each cell is one of several distinct cell types, e.g., CD8$^+$ T cell, and that each cell type is capable of presenting a set of membrane-bound receptor proteins on its surface with the set of receptors associated with different cell types potentially overlapping. We additionally assume that each receptor has a unique cognate secreted ligand protein that can bind to it and that each ligand is produced as a consequence of one or more receptor signaling pathways, i.e., binding of a given receptor by its associated ligand will trigger production of other ligands by the cell. 

Given this simple ligand/receptor signaling model and the distribution of cells within a tissue, a key question is to estimate the steady-state activity of each receptor signaling pathway. One approach for answering that question creates a multilayer network with one layer per receptor protein with the activity of a specific receptor signaling pathway in a given cell represented by the centrality of the associated node. Although we can assume that the adjacency matrices for all receptor layers are identical, a more realistic model would vary the edge weights (potentially with thresholding) based on the dispersion properties of each cognate ligand. Simply computing the eigenvector centrality for each layer, however, does not yield the appropriate answer since the importance of adjacent cells in a given receptor layer reflects that activity of that receptor, which most likely does not impact the activity of that same receptor in other cells, i.e., binding of a given receptor does in general result in secretion of the associated ligand. Instead, one wants to use the importance of cells in the layers corresponding to receptors that produce the cognate ligand. This type of inter-layer dependency structure is exactly what our proposed model supports. We believe that a more general class of systems biology questions may map to similar interdependent multilayer networks.

Our future work in this area includes exploring the theoretical properties of our multilayer network model and interleaved power iteration algorithm (e.g., iteration convergence), characterizing the computational performance on a range of simulated and real networks, performing a comparative evaluation against other multilayer centrality approaches, and applying our technique to study the example cell signaling problem using tissue imaging and genomic profiling data.

\section*{Acknowledgments}

This work was funded by National Institutes of Health grants R35GM146586, R21CA253408, P20GM130454 and P30CA023108.
We would like to thank Peter Bucha and James O'Malley for the helpful discussions and feedback.
We would also like to acknowledge the supportive environment at the Geisel School of Medicine at Dartmouth where this research was performed. 

\bibliographystyle{unsrt}
\bibliography{ConstrainedMultilayerCentrality.bib}

\end{document}